\pdfoutput=1
\documentclass[twocolumn]{aastex63}
\usepackage{hyperref}
\usepackage{amsmath}
\usepackage[switch]{lineno} 
\usepackage{listings}
\usepackage{xcolor}

\definecolor{codegreen}{rgb}{0,0.6,0}
\definecolor{codegray}{rgb}{0.5,0.5,0.5}
\definecolor{codepurple}{rgb}{0.58,0,0.82}
\definecolor{backcolour}{rgb}{0.95,0.95,0.92}

\lstdefinestyle{mystyle}{
    backgroundcolor=\color{backcolour},   
    commentstyle=\color{codegreen},
    keywordstyle=\color{magenta},
    numberstyle=\tiny\color{codegray},
    stringstyle=\color{codepurple},
    basicstyle=\ttfamily\footnotesize,
    breakatwhitespace=false,         
    breaklines=true,                 
    captionpos=b,                    
    keepspaces=true,                 
    numbers=left,                    
    numbersep=5pt,                  
    showspaces=false,                
    showstringspaces=false,
    showtabs=false,                  
    tabsize=2
}

\newcommand{\tl}{\texttt{TESS\_localize}}

\submitjournal{AAS Journals}

\shorttitle{Localizing Blended Variable Stars in TESS}
\shortauthors{Higgins \& Bell}

\begin{document}

\title{Localizing Sources of Variability in Crowded TESS Photometry}

\correspondingauthor{Michael Higgins}
\email{michael.higgins@duke.edu}

\author{Michael E.\ Higgins}
\affil{Department of Physics, Duke University, Durham, NC-27708, USA}

\author[0000-0002-0656-032X]{Keaton J.\ Bell}
\altaffiliation{NSF Astronomy and Astrophysics Postdoctoral Fellow}
\altaffiliation{Current address: Department of Physics, Queens College, City University of New York, Flushing, NY-11367, USA}
\affil{DIRAC Institute, Department of Astronomy, University of Washington, Seattle, WA-98195, USA}

\begin{abstract}
    The Transiting Exoplanet Survey Satellite (TESS) has an exceptionally large plate scale of 21\arcsec\,px$^{-1}$, causing most TESS light curves to record the blended light of multiple stars.  This creates a danger of misattributing variability observed by TESS to the wrong source, which would invalidate any analysis. We developed a method that can localize the origin of variability on the sky to better than one fifth of a pixel. Given measured frequencies of variability (e.g., from periodogram analysis), we show that the best-fit sinusoid amplitudes to raw light curves extracted from each pixel are distributed the same as light from the variable source. The primary assumption of this method is that other nearby stars are not variable at the same frequencies. Essentially, we are using the high frequency resolution of TESS to overcome limitations from its low spatial resolution.  We have implemented our method in an open source Python package, \tl\ (\href{https://github.com/Higgins00/TESS-Localize}{github.com/Higgins00/TESS-Localize}), that determines the location of a variable source on the sky and the most likely \textit{Gaia} source given TESS pixel data and a set of observed frequencies of variability. Our method utilizes TESS Pixel Response Function models, and we characterize systematics in the residuals of fitting these models to data. We find that even stars greater than three pixels outside a photometric aperture can produce significant contaminant signals in the extracted light curves. Given the ubiquity of source blending in TESS light curves, verifying the source of observed variability should be a standard step in TESS analyses.

\end{abstract}


\section{Introduction}
    The Transiting Exoplanet Survey Satellite (TESS) is executing a thorough census of photometric variability over 85\% of the sky \citep{TESS}. It collects continuous images in a series of pointings (sectors), visiting each of these overlapping fields for approximately 27 days. During its first two years of operations, TESS obtained full-frame images (FFIs) every 30\,min, as well as 2-min cadence subframe images around $\approx 20{,}000$ bright stars and other targets of specific interest per sector \citep{2021PASP..133i5002F}. The current extended mission has increased the FFI rate to 10\,min, added a 20-sec cadence, and increased the number of short-cadence targets in each sector.  The records of stellar variability obtained during these extensive photometric campaigns achieve frequency resolutions of $\approx\,0.43\,\mu$Hz or finer, and they are not confused by aliasing from daytime gaps in the data that affect ground-based surveys.
    
    What TESS achieves in frequency resolution, it lacks in spatial resolution. TESS has an exceptionally large plate scale of 21\arcsec\,px$^{-1}$. As a result, multiple stars often contribute light to the same detector pixels, and it can be difficult to identify which star is the source of recorded variability. The source blending problem varies with direction, with upwards of dozens of stars brighter than magnitude 18 contributing light to most pixels near the galactic plane. The amount of contamination for each target is approximated by the {\sc crowdsap} value in the TESS FITS file headers, which gives the ratio of target star flux to total flux in the photometric aperture. When analysing TESS light curves, there is considerable risk of attributing detected variability to the wrong source, leading to erroneous conclusions.
    
    Even with 4\arcsec\ pixels, blending was a concern for analyses of data from the \textit{Kepler} spacecraft.  \textit{Kepler} \citep{Kepler} can be considered a predecessor of TESS that also obtained continuous light curves of stars primarily to detect  exoplanet transits, though in a field of view 400 times smaller. \textit{Kepler} observed over 500,000 stars from which a list of \textit{Kepler} objects of interest (KOIs) that had transit-like signals was compiled. Nearly half of these KOIs turned out to be false positives caused by contamination from eclipsing binaries \citep{Bryson2013}.  \citet{Colman2017} demonstrated another example of blended starlight complicating analyses of \textit{Kepler} light curves, showing that anomalous peaks in the periodograms of red giant stars could indicate a companion orbiting within the convective giant envelope, while in many cases such signals likely originate from chance alignments with other variable sources in the field.
    
    Since \textit{Kepler} and TESS are motivated by the search for transiting exoplanets, many tools have been developed to vet candidate transit signals. Shallow planet transits can be mimicked by contaminating light from eclipsing binary systems. \citet{Giacalone_2020} review the history of software tools for vetting candidate transits that test the data against transit and blended-binary models, and introduce the {\tt Triceratops} package that utilizes the \textit{Gaia} DR2 star catalog to model the possible contamination from many stars that is especially relevant for TESS. The software tool {\tt LATTE} provides an interactive interface for investigating potential contaminant or systematic origins of candidate transit signals in TESS \citep{Eisner2020}. One such diagnostic is to test whether the centroid of light from a target source moves during transits or eclipses, which could indicate that the transited star is spatially offset from the target \citep{Bryson2013,Vetting}.
    
    Time series photometry from TESS is invaluable for studying many types of brightness variability of astronomical objects.
    Some methods have been developed to address the challenges of source blending in general. \citet{2018AJ....156..132O} presented a data reduction pipeline to remove blended light from photometry through a method called difference imaging analysis (DIA) where a high signal-to-noise image stack is subtracted from a reference frame to remove any non-variable signal in the photometry. The DIA method is useful when looking for transient events. The method used in \citet{Colman2017} to identify whether signals originate from their target stars or nearby contaminating sources was to compute a periodogram for light curves extracted from each pixel in a \textit{Kepler} ``postage stamp''\footnote{The term "postage stamp" is commonly used to refer to the subset of pixels read out around targets of interest to construct a target pixel file.} and identify where the corresponding peaks appear to be centered in the pixels. This process was done by visual inspection to get a general idea of the location of the source of the anomalous peaks in the periodograms and spatially resolve them. \citet{Hedges2021} developed a method that can extract individual, deblended light curves for sources that are separated by at least one pixel and differ in brightness by more than a magnitude. 
    In many cases, potential source confusion can be resolved by comparing to additional archival or follow-up time series photometry that is spatially resolved for the various sources in the field \citep[e.g.,][]{Collins2018}.
    To aid in disentangling variability of blended sources in northern TESS Cycle 2 fields, the Zwicky Transient Facility conducted a nightly, contemporaneous photometric survey at high angular resolution that can be compared to TESS data to match variability in many cases \citep{TESS-ZTF}.  
    
	This paper presents a new method to solve the persisting problem of spatially resolving signal sources in low-resolution time series survey photometry. Our strategy for localizing where variability is originating from in the pixel data is to fit the spatial distribution of signal amplitudes for frequencies of interest. We recommend verifying the source of variability for all analyses of TESS data where significant frequencies of variation can be measured. We are releasing a Python package, \tl\ at \href{https://github.com/Higgins00/TESS-Localize}{github.com/Higgins00/TESS-Localize}, that can fit the sky location of observed variability in the TESS pixel data. We detail the method in Section~\ref{sec:methods}, validate its performance with simulations in Section~\ref{sec:simulations}, explore some case studies with real TESS data in Section~\ref{sec:casestudies}, and provide practical guidance for users in Section~\ref{sec:conclusions}.

\section{Methods}  \label{sec:methods}

    In general, TESS pixels record the combined light contributions from many sources. If we detect variability in an extracted light curve, it is not immediately clear which source the variability originates from. We have developed a method for localizing the source of detected variability in the TESS pixel data based on the fact that the (unnormalized) amplitude of variation in each pixel is proportional to the flux contribution from the variable source. Essentially, we are able to resolve the variable source in frequency space for each pixel. We develop the concepts and assumptions behind the method before detailing its implementation in the Python package that we are releasing as an open-source research tool.
    
    \subsection{Overview}
    
    We will consider our source of interest to be some theoretical variable star with average total flux $F$ measured by the detector. 
    This flux is distributed on the detector following the TESS pixel response function (PRF\footnote{\url{https://heasarc.gsfc.nasa.gov/docs/tess/observing-technical.html##point-spread-function}}; \citealt{Bryson2010}). The PRF represents the point spread function convolved with average pointing jitter during an exposure, as recorded by each pixel.
    We indicate the fraction of the source flux as distributed by the PRF measured at  the pixel column $i$ and row $j$ as $P_{i,j}$, assuming for now that pointing is steady during the observations. 
    The flux in each pixel from the variable source is then $F*P_{i,j}$.
    This light will be blended with a potentially complicated distribution of background light from other nearby sources, and we represent the average fractional ratio of the  flux from the source star to the total flux in each pixel as $C_{i,j}$.\footnote{Analogous to the {\sc CROWDSAP} header value that approximates crowding in the Science Processing Operations Center pipeline apertures.} Dividing by this, we find the average total flux measured from all sources in each pixel to be $F*P_{i,j}/C_{i,j}$. 
    
    The primary assumption underlying our method is that the other nearby stars do not exhibit significant variations at frequencies that are unresolved from those of our source.
    Here, nearby means contributing light to the target pixel files (TPFs) that are read out around every short-cadence target, or to the subframe cutouts acquired from the full frame images with {\tt TESSCut} \citep{TESSCut}. For simplicity, we will refer to both as TPFs in this work.
    Putting this assumption another way, the only appreciable effect that other stars have on the periodograms of light curves extracted from a TPF is on the measurement noise at the frequencies specified.
    This assumption will be satisfied in most cases, owing to the high frequency resolution achieved by the TESS data of $\approx0.43\,\mu$Hz, as well as the minimal aliasing. For 2-minute cadence data, for instance, this resolution corresponds to over 10{,}000 effectively independent frequency bins below the Nyquist frequency of 4467\,$\mu$Hz. It is unlikely that the frequencies of variation of blended background stars will happen to fall within the same frequency bins as our target source variability. We provide practical guidance for validating this assumption in Section~\ref{sec:conclusions}.
    
    Suppose the target star of interest varies sinusoidally with frequency $\nu$ and relative (fractional) amplitude $a$. Since we assume that only the source exhibits significant power at frequency $\nu$, we can fit a model for how the flux varies \textit{at this frequency only} in each pixel
    \begin{equation}
    \begin{aligned}
    F_{{\rm \nu}(i,j)}(t) &= F \times P_{i,j} (a\sin{(2\pi\nu t+\phi)} + 1/C_{i,j}) \\
    & = A_{i,j} \sin{(2\pi\nu t+\phi)} + \langle {\rm Flux_{i,j}}\rangle,
    \end{aligned}
    \end{equation}
    where $\phi$ is a phase term shared across pixels. This assumes that the detected flux is linearly proportional to the incident flux, which will not be true for bright stars that saturate pixels \citep[e.g.,][]{2017MNRAS.471.2882W}.
    This is essentially the Fourier component at frequency $\nu$, plus an offset for the average flux in the pixel.
    Since $F$ and $a$ are constant values, the fitted amplitude $A_{i,j}$ is proportional to $P_{i,j}$, the fraction of light from the variable source in pixel $(i,j)$. Therefore, if we fit this model for every pixel in the TPF, we can empirically recover how the light from the variable source is distributed on the detector. In practice, the TESS pointing drifts during a sector by $\sim0.01$\,pix, so the average amplitude distribution will be slightly broader than the per-frame PRF. Fitting a PRF model to this average amplitude distribution yields an estimate of the location of the variable source on the sky. 
    
    This localization procedure can be improved by fitting a common location for multiple frequencies of variation, assuming that they all arise from the same target. General non-sinusoidal variations can be represented as a sum of sine waves (harmonics for periodic variability).
    
    \subsection{Implementation}
    
\begin{figure}
        \centering
        \includegraphics[width=\columnwidth]{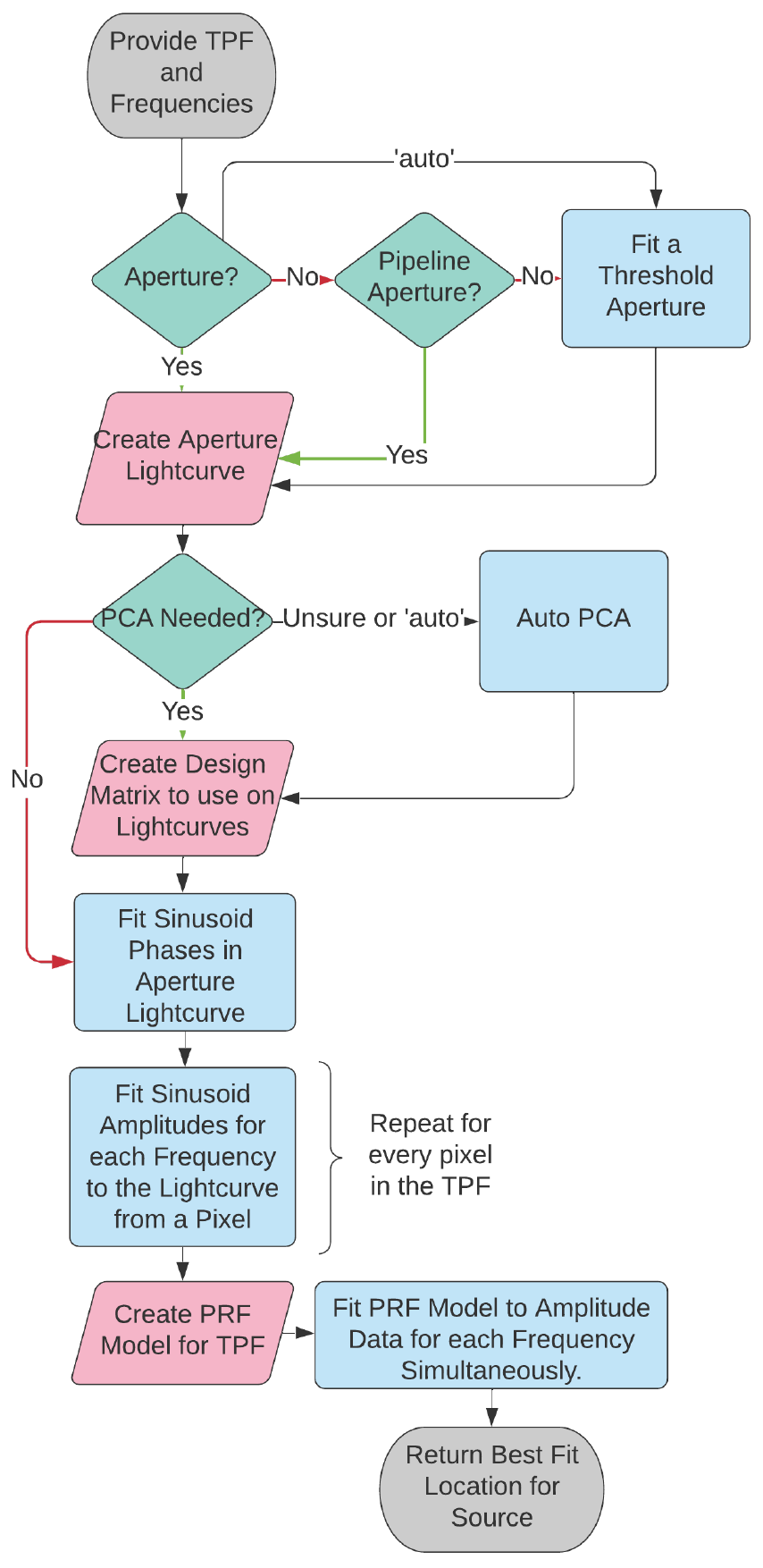}
        \caption{Flow chart of the \tl\ method described in this work. The first grey box and the green diamonds are the only parts of the software that require user input.}
        \label{fig:flowchart}
    \end{figure}

    In practice, we anticipate that users will have identified frequencies of variability from a periodogram analysis of a light curve, and they then wish to verify where on the detector those variations arise from. Our implementation of this signal localization method in the Python package \tl\ requires a user to provide a list of frequencies, TPF data (as a \texttt{lightkurve.TargetPixelFile} object), and optionally the aperture used to extract the light curve that the frequencies were measured from.  If an aperture is not provided, the TPF pipeline aperture will be used by default. There is also an option that attempts to automatically choose an appropriate aperture containing the highest Lomb-Scargle periodogram power at the provided frequencies. The flow of how the software proceeds based upon user input is presented in Figure \ref{fig:flowchart}.
    
    With real data, systematic trends may be present in the flux measured by various pixels that could mimic signals at the frequencies of interest. These could be caused by scattered light or spacecraft motion causing starlight to drift across different pixels. Before fitting, \tl\ provides the option to identify trends that are most common among pixels outside the provided aperture with principal component analysis (PCA) using the {\tt RegressionCorrector} tools from the {\tt lightkurve} package \citep{lightkurve}.\footnote{See the {\tt lightkurve} tutorial on {\tt +RegressionCorrector} at \url{docs.lightkurve.org/tutorials/2-creating-light-curves/2-3-removing-scattered-light-using-regressioncorrector.html}}
    These PCA components can be fit to and subtracted from all of the light curves that \tl\ extracts from the pixels. It is, however, important to ensure that these PCA trends do not include the signals that you aim to localize, or the signals will be removed from the data. The \tl\ user can inspect and choose which PCA components to include in their analysis by using the \texttt{TESS\_localize.PCA()} function. \tl\ can optionally attempt to determine the optimal number of PCA components to use by selecting the strongest trends that do not appear to contain significant power at any of the frequencies in the frequency list input by the user, this is determined by going from the most to the least principal component until a component is found to have power at 5 times the median power for any of the frequencies provided. This method of picking the optimal number of PCA components primarily focuses on not removing signals of stellar variability from the light curves and is a crude estimation that the careful user is encouraged to scrutinize.
    
    Intrinsic variability from any individual source will arrive with the same phase across the image.  \tl\ performs an initial fit of a sum-of-sinusoids model to the light curve extracted from the aperture, which should provide a high signal-to-noise measurement of signal phases. The first fit fixes amplitudes based on the Lomb-Scargle periodogram amplitudes evaluated at the provided frequencies, fixes the average flux to the mean measured flux, and brute-force samples phase in 20 evenly sampled steps to obtain good starting phase values for further optimization.  Then a second, non-linear least-squares fit is performed where phases, amplitudes, and mean flux are all free to vary.  These fits are all performed using the implementation of the Levenberg–Marquardt algorithm via the Python package \verb+lmfit+ \citep{lmfit}.
    
    With the signal phases measured, \tl\ moves on to fit the signal amplitudes from the unnormalized light curves extracted from each pixel in the TPF.
    \tl\ then performs least-squares fits of our (multi-)sinusoid model to the light curve from each pixel with \verb+lmfit+, with frequencies fixed to the input values, and phases fixed to the best-fit values to the aperture light curve from the previous step. The uncertainty on the fitted amplitudes are expected to be $\sigma_a = \sqrt{2/N}\sigma_F$, where $\sigma_F$ is the typical light-curve flux error, and $N$ is the number of points in the light curve \citep{Montgomery1999}. Since $\sigma_F \approx \sqrt{F}$, pixels with more flux contamination will yield noisier measurements of the signal amplitudes. This is accounted for when least-squares fitting sinusoids to the light curves by weighting by the reciprocal of the flux uncertainties.
    
    With amplitudes measured for each frequency in each pixel, \tl\ finally fits the common location where this variability arises from. We created a Python package called {\tt TESS\_PRF}\footnote{\url{https://github.com/keatonb/TESS_PRF}} that interpolates the TESS PRF files created by the TESS Science Processing Operations Center \citep[SPOC;]{2016SPIE.9913E..3EJ} available on the Mikulski Archive for Space Telescopes (MAST) for a given location on the detector, which we describe in Appendix~\ref{app:tessprf}. \tl\ simultaneously fits a PRF model to the location in the TPF from which all provided frequencies are most consistent with originating, scaled to different amplitudes for each signal. With \verb+lmfit+, the program minimizes the absolute difference between the pixel-by-pixel amplitudes from the previous step and the PRF model, divided by the amplitude fit uncertainties.
    
    Differential velocity aberration causes TESS pointing to drift slightly during a sector, as recorded by the TPF data columns `POS\_CORR1' and `POS\_CORR2,' which give column and row pixel offsets relative to WCS metadata. The typical standard deviation of this pointing variation is $\approx 0.02$\,pix, which will broaden the amplitude heatmap slightly, but not enough to affect our fitting significantly. \tl\ adjusts the best-fit source location to account for the average pointing offset for that TPF. This pointing information is not currently preserved in TPF-like data produced with {\tt TESSCut},\footnote{A fix for this issue has been requested:\\ \url{https://github.com/spacetelescope/astrocut/issues/58}} so results from {\tt TESSCut} images may have inaccuracies of-order 0.01\,pix.
    
    The \tl\ program returns the optimized column and row location within the TPF where the observed variability is most likely located, with intrinsic uncertainties.   The code adopts the FITS WCS standard that integer pixel locations represent the center of pixels \citep{fitswcs}. For each signal in the input frequency list, a ``heatmap'' can be displayed with the {\tt plot} function showing in each pixel the signal amplitude, amplitude error, signal-to-noise ratio, or the amplitude - model residuals.  The overall fit report including fit statistics such as (reduced) chi-square is stored as the ``result'' parameter. Best-fit parameters for the amplitude of each signal are returned with the prefix ``height.''
    
    \tl\ can optionally query the \textit{Gaia} Archive for the locations of stars in the field, ranking them by the relative probability that their locations are consistent with the source of variability \citep{2018}. We propagate star positions by proper motions from the reference epoch to the TESS epoch using \textit{Gaia} proper motions.
    Probabilities are calculated using the extrinsic error model discussed in Section~\ref{sec:errormodel}. The extrinsic error accounts for systematic errors that arise when fitting a PRF model to real TESS data. 
    Two metrics are provided to help the user interpret the fit results: p-values and relative likelihoods. A ``p-value'' represents the fraction of possible locations around each star with lower likelihood (as evaluated in the error model) than the actual fit location. These numbers could be used to reject the hypothesis that the position of each star is consistent with the fit location if the p-value is below an acceptable value.
    The ``relative likelihood'' values reported give the relative probabilities of each star location corresponding to the fit location, normalized to 1.0 under the assumption that the localization corresponds to one of the considered sources. This assumption should be supported by a p-value that exceeds an appropriate threshold for your project.

\section{Simulations}\label{sec:simulations}
    To assess the performance of our method under ideal conditions, we simulated TESS-like pixel data for four scenarios. These tests provide empirical evidence that our implementation of this method recovers accurate variable source positions, and that the intrinsic uncertainties given by the fitting procedure closely match the residuals. The four situations we simulated are as follows:
\begin{enumerate}
    \item An isolated star that is variable with different amplitudes relative to noise.
    \item A star that is at varying distances to the edge of a TPF.
    \item Two blended stars at varying separations from each other.
    \item Two blended variable stars with some variability frequencies in common.
\end{enumerate}

For these simulations, we approximate TESS TPF data by generating sequences of 11x11 pixel images every 2\,min for a duration of 27 days. Light from each star is distributed as a Gaussian across the pixels with a scale factor (standard deviation) of 0.7\,pix.  These Gaussians are sampled nine times finer than the plate scale in each dimension and integrated over the pixel locations to approximate a TESS PRF. In the first three simulations, the star that is variable is only variable at one signal that has a period of 10 days.  Our method will perform better for sources that vary with multiple frequencies, which boosts the effective signal-to-noise for our fitting procedure. A 5-pixel cross-shaped aperture centered at the simulated star is used to determine the signal-to-noise ratio. The noise is simulated by adding normally distributed independent random values scaled to the square root of the flux in each pixel of each image.

\subsection{Case 1: Star Variability Signal-to-Noise}
\label{sec:case1}

        \begin{figure}[!h]
        \centering
        
            \includegraphics[width=\columnwidth]{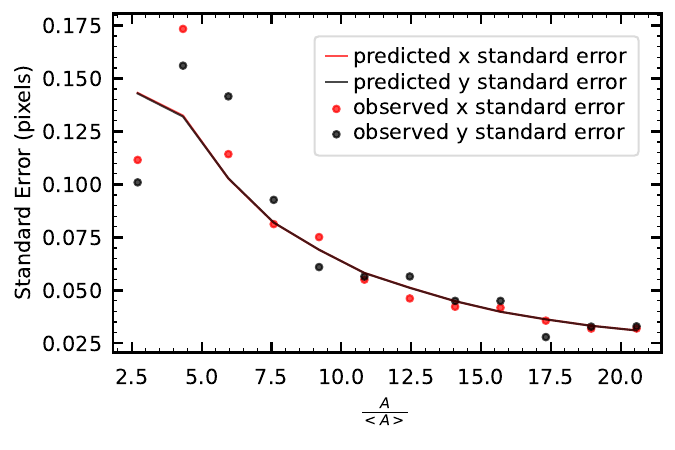}
            \caption{Predicted standard errors from the fitting procedure (solid lines) and the observed standard deviation of the fit residuals (points) as a function of signal-to-noise in the aperture.  The signal-to-noise is computed as the signal amplitude divided by mean amplitude in the periodogram of the light curve.  We ran 1000 simulations for each signal-to-noise value tested.}
            \label{fig:ss5s}
        \end{figure}
        
        \begin{figure}[!h]
        \centering
            \includegraphics[width=\columnwidth]{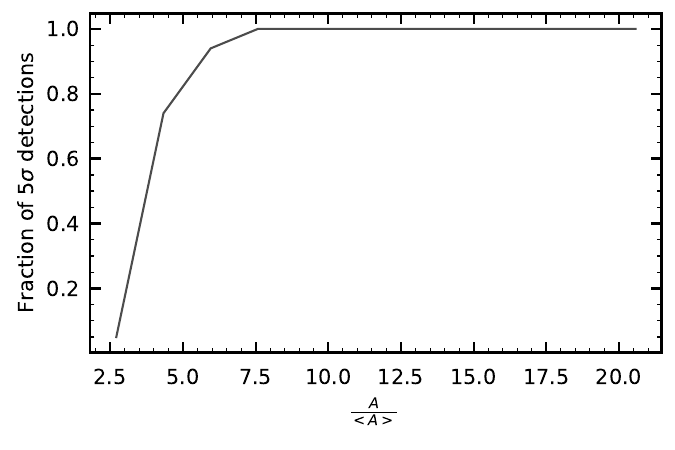}
            \caption{Fraction of source star fits with significant detections at the signal-to-noise levels presented in Figure~\ref{fig:ss5s}. Fits with relative amplitude errors less than 20\% are considered significant.}
            \label{fig:ssse}
        \end{figure}

    We simulated a star 1000 times in the center of the TPF for multiple signal-to-noise ratios.  As a metric for signal-to-noise, we define A/$\langle \mathrm{A} \rangle$ as the amplitude of our signal divided by the mean amplitude in the periodogram of the light curve extracted from a 5-pixel cross-shaped aperture centered about the true location of the simulated star. For context, simulating 10,000 light curves shows that a signal with A/$\langle \mathrm{A} \rangle$ greater than 4.4 has less than a 0.1\% probability of being caused by pure noise \citep[false alarm probability; see also][]{2021AcA....71..113B}. We applied our source localization method to all 1000 simulated stars per A/$\langle \mathrm{A} \rangle$ to obtain the best-fit location with estimated uncertainties. 
    
    As a criterion for a significant detection, we keep only the results that have a fractional error on the fit of the ``height'' of the Gaussian-distributed signal less than 20\%, corresponding to a 5$\sigma$ significance. The distribution of the locations fit for these stars is approximately Gaussian. Figure \ref{fig:ss5s} displays the average standard error reported by our fitting procedure as a function of A/$\langle \mathrm{A} \rangle$ as solid lines. The standard deviations of the residuals between simulated and fitted position are plotted as points for each signal-to-noise level, and they follow the reported uncertainties from the fits to within a few percent. Figure \ref{fig:ssse} shows the fraction of simulations that yielded a 5$\sigma$ detection for different signal-to-noise ratios. Our method successfully fits signals detected to a significance of A/$\langle \mathrm{A} \rangle \geq 7.5$ every time, with intrinsic errors less than 0.1\,pix in each direction.
 
 This result presents an opportunity to check that the signals in the extracted light curve are strong enough for reliable localization. For a light curve with Gaussian distributed noise, the noise in the periodogram follows a Chi distribution with two degrees of freedom ($\chi_2$). For $N$ time series observations with a standard deviation of the noise $\sigma_F$, the expected mean of the $\chi_2$-distribution is 
 $\langle \mathrm{\bar{A}} \rangle = \Gamma(3/2)\sigma_F\sqrt{4/N} \approx 1.253\sigma_F\sqrt{2/N}$. Rearranging from \citet{Montgomery1999}, the fit uncertainty on the phase of a  sinusoidal signal with amplitude $A$ is expected to be $\sigma_\phi = \sigma_F\sqrt{2/N}/A$. We use these expressions to convert the reliable recovery threshold of A/$\langle \mathrm{\bar{A}} \rangle \gtrsim 7.5$ to an approximate upper limit on the phase uncertainty $\sigma_\phi \lesssim 0.1$ radians. We include a check at the phase-fitting stage of \tl\ to warn the user if large phase uncertainties suggest the signals may be too weak in the aperture to be localized.

\subsection{Case 2: Star Location}
    \begin{figure}
        \centering
        \includegraphics[width=\columnwidth]{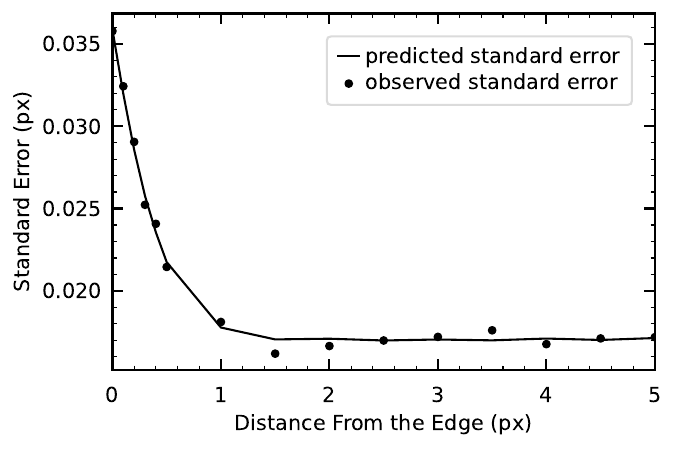}
        \caption{Same as Figure~\ref{fig:ss5s}, but as a function of the variable star distance from the edge of the TPF.}
        \label{fig:locationsim}
    \end{figure}

    To test whether the precision of our method degrades when light from the variable star is not entirely contained within the TPF, we simulated a star 1000 times each at locations with various distances from the edge of the image in the Y-axis, but centered in the X-axis. We found that a star with a A/$\langle \mathrm{A} \rangle > 7$  could be successfully fit to the edge of a TPF with fit residuals that agree to within a tenth of the reported errors. Figure \ref{fig:locationsim} shows the predicted and observed uncertainties for varying source distances from the edge of the TPF.
    Fit locations become less precise as light from the source falls off the edge of the TPF, but this is accurately reflected in the reported intrinsic error. The \tl\ code raises a warning if the fit indicates that most of the flux is located outside the TPF. A more precise localization might be achieved using {\tt TESSCut} \citep{TESSCut} to retrieve pixel data encompassing more light from the source, although this is only available for long-cadence data.

\subsection{Case 3: Two Blended Stars}

    \begin{figure}
        \centering
        \includegraphics[width=\columnwidth]{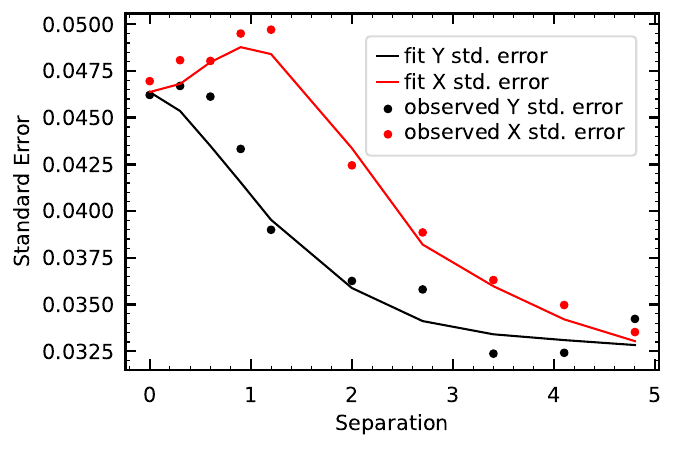}
        \caption{Same as Figure~\ref{fig:ss5s} for each the X and Y directions as a function of the separation between the source star and the non-source star.}
        \label{fig:tsse}
    \end{figure}
    
    To determine the effectiveness of our method where light from multiple sources is blended in the same pixels, we introduced a second star with constant brightness at different distances in the X direction from our target star. This second star was set to have the same average flux as the variable star. The variable star by itself has a flux of 1000 and a variability of 3\% resulting in a A/$\langle \mathrm{A} \rangle$ by itself of approximately 16, which is sufficient for our code to obtain a $>5\sigma$ detection for all trials. For each separation from the central variable star, we simulated 1000 data sets.
    
    The plot of the typical reported and observed uncertainties in the best-fit column and row for different source separations in Figure~\ref{fig:tsse} shows three distinct features. First, column and row uncertainties match when the stars are simulated at the same position or are completely separated. Second, the error in column position initially increases with separation as the combined starlight is elongated along this dimension, and the additional contaminant flux increases the fitting uncertainty on signal amplitudes. Finally, the overall precision improves as separation increases since the A/$\langle \mathrm{A} \rangle$ in the target aperture increases as contamination decreases.  Despite these effects, the uncertainties reported by our fitting procedure closely match the typical residuals for these simulations.
    
\subsection{Case 4: Two Blended Multi-variable Stars}
        
    A final simulation test was run to understand the effectiveness of our method when multiple stars are blended and share some of the same frequencies of variability. We do not expect that localizations of signal frequencies shared between multiple stars in the TPF will be accurate, as this violates the main assumption of the method. We simulated two variable stars, each with eight frequencies of variability where four of these frequencies were shared between the two stars. All signals were of equal amplitude. In each iteration of the simulation, the stars shared the same total flux but increased in separation from each other. For each separation, we used 800 realizations to compute the averages for our analysis.
    
\begin{figure}
        \centering
        \includegraphics[width=\columnwidth]{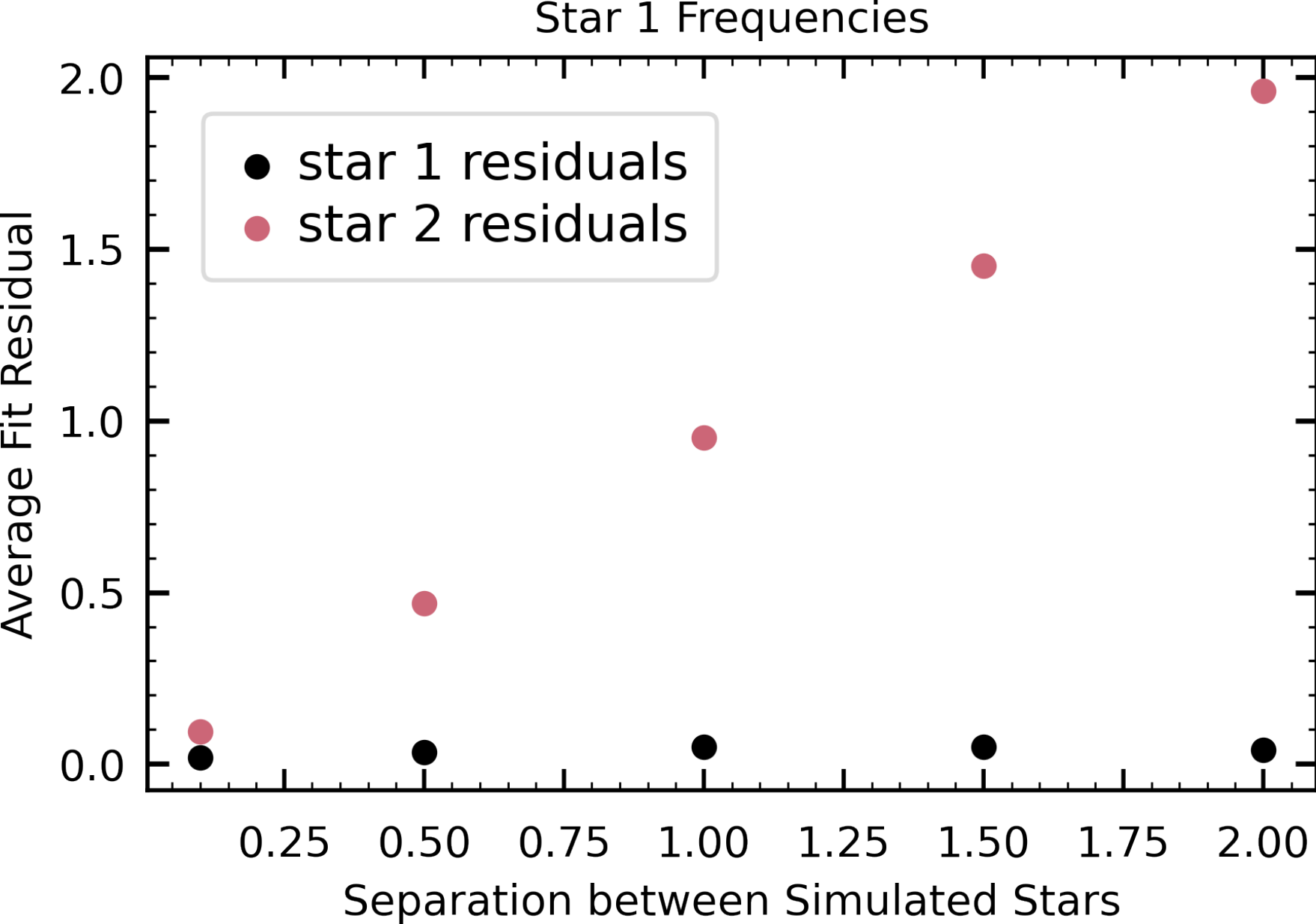}
        \caption{Average distance between the fitted location and the true location of two stars as a function of their simulated separation (in pixels) for the 4 frequencies unique to star 1. Result consistently fit to the correct location of star 1.}
        \label{fig:sharedfreqstars_star1}
    \end{figure}

    When fitting the four signals that were unique to each star, our method was capable of fitting the locations accurately and consistent with the reported errors. Figure~\ref{fig:sharedfreqstars_star1} shows the distance between the fit location and each star location when fitting the frequencies unique to star 1; despite having other frequencies in common, the fit is always consistent with the position of star 1, while fit distance from star 2 matches the star-star separation. Results are similar for the frequencies unique to star 2. The reported uncertainties on position when fitting four unique frequencies are a factor of 2 smaller than when using just one frequency.

    \begin{figure}
        \centering
        \includegraphics[width=\columnwidth]{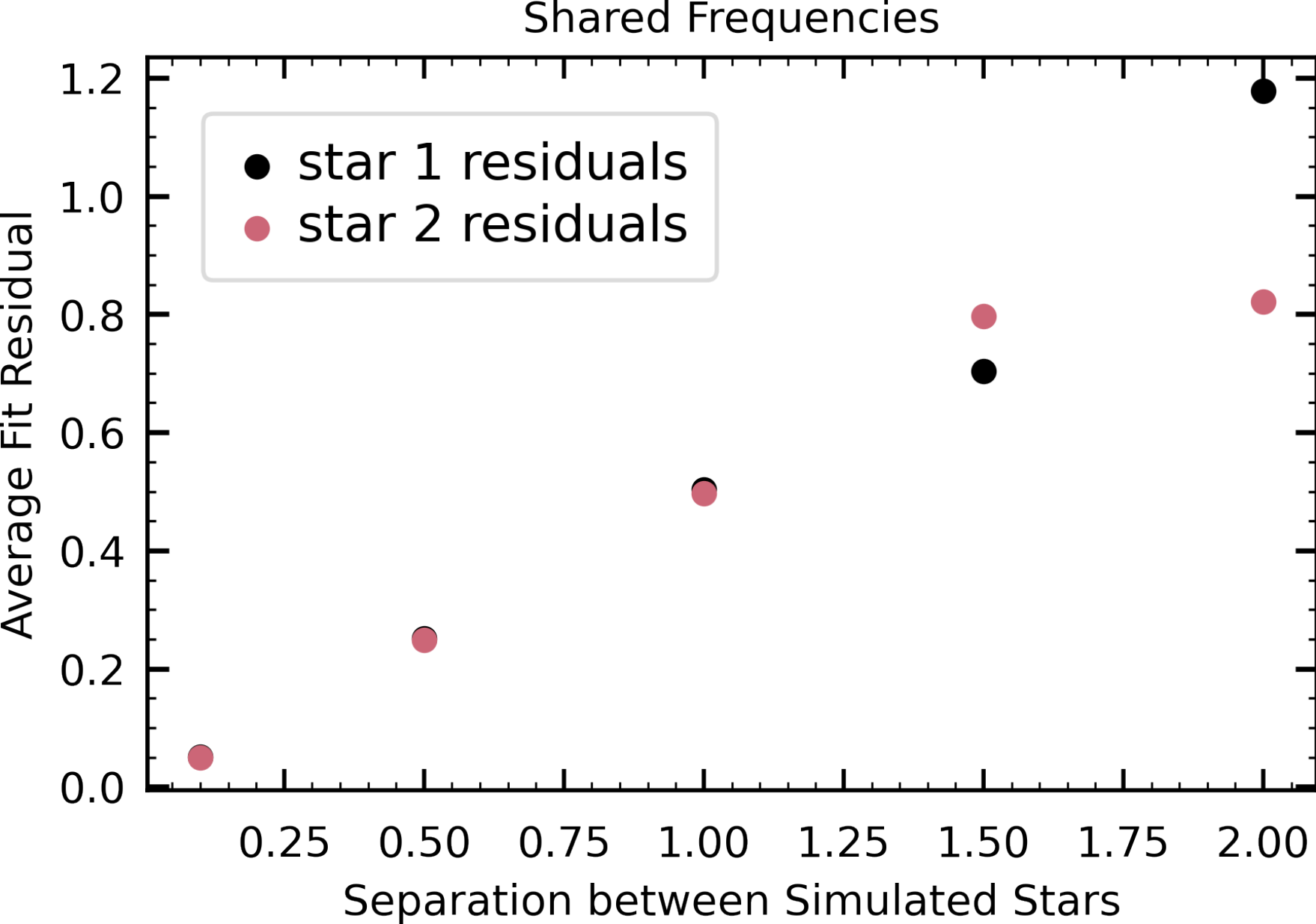}
        \caption{Same as Figure~\ref{fig:sharedfreqstars_star1}, but for the 4 frequencies that are shared between the two stars.}
        \label{fig:sharedfreqstars_shared}
    \end{figure}

    \begin{figure}
        \centering
        \includegraphics[width=.8\columnwidth]{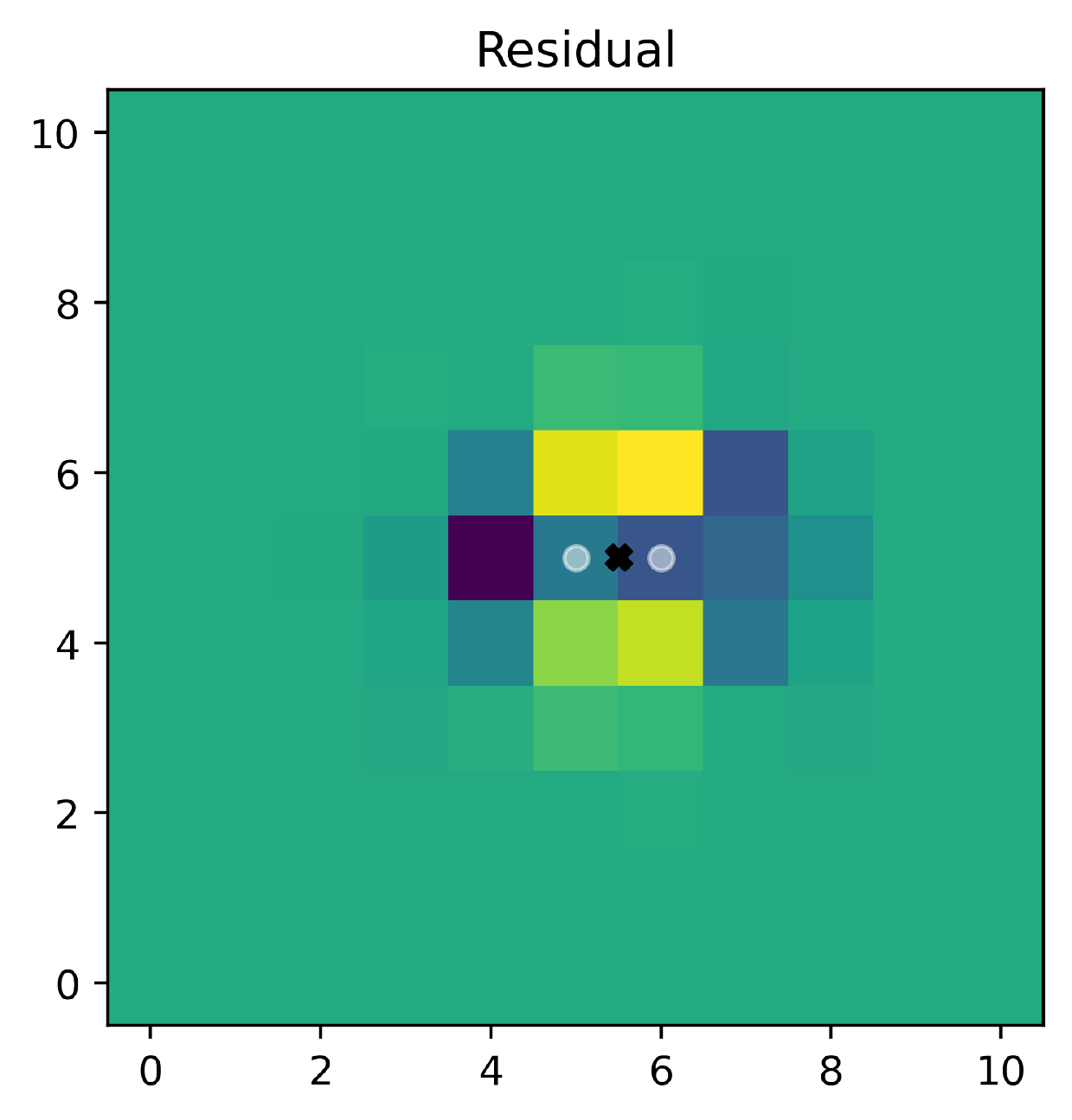}
        \caption{Residual between amplitude heatmap and the best-fit single-star model for two simulated stars with shared frequencies that are in phase with each other. The black 'X' marks the fit location, and the grey circles mark the locations of the two stars. Model values are greater than the measured amplitude distribution in the bright yellow pixels, and less than the amplitudes in the dark blue pixels.}
        \label{fig:resinphase}
    \end{figure}    

    \begin{figure}
        \centering
        \includegraphics[width=.8\columnwidth]{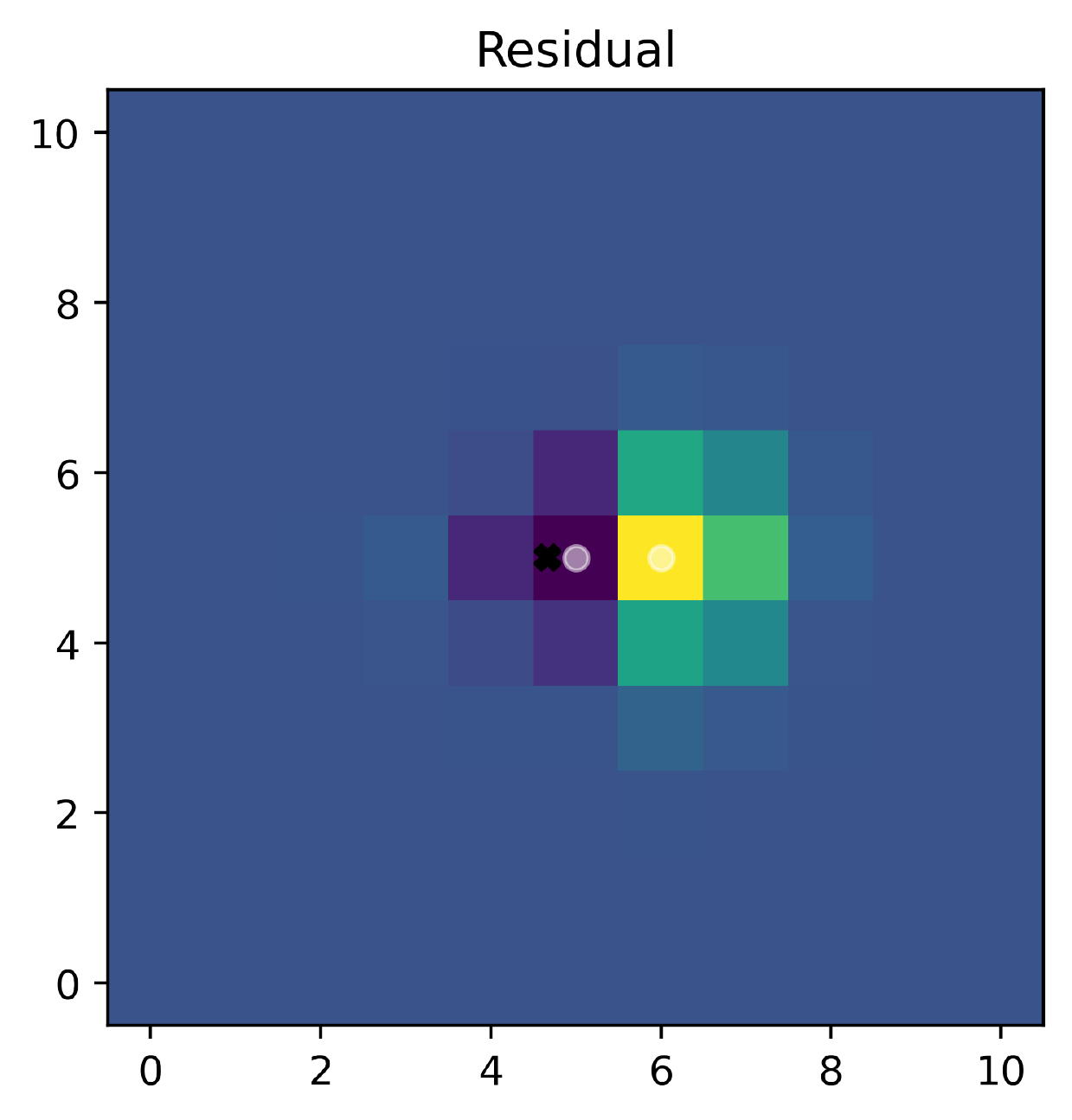}
        \caption{Same as Figure~\ref{fig:resinphase} for shared frequencies that are $\pi$ out of phase with each other.}
        \label{fig:resoutphase}
    \end{figure}

    When fitting a source location for the four frequencies the stars had in common, our code reports a fit position that is inconsistent with the true location of either star. Figure~\ref{fig:sharedfreqstars_shared} shows the results from attempting to fit the location of the shared frequencies. On average, the results are equidistant from both stars in ranges where they are significantly blended (within about 1 pixel in separation), then in the ranges where they are adequately separated the fit residual diverges as the fit is attracted more to one star or the other. The effect depends on the relative phases of the shared signal frequency: when the signals are in phase, the fit location is equidistant between the two blended stars; when the signals are perfectly out of phase, there is a repelling effect as a result of one star having a negative amplitude at the shared model phase. 
    One may be able to identify cases where two stars in a TPF share signals by inspecting the pixel-by-pixel residuals for each frequency. Figures~\ref{fig:resinphase} and \ref{fig:resoutphase} show examples of what the residuals look like for simulated stars separated by 1 pixel with the same or opposite phases, respectively. Regardless of phase difference, there is considerable structure in the residuals. Given the high frequency resolution of TESS, it is likely that issues arising from stellar sources sharing significant power at the exact same frequencies will be uncommon.

\section{Demonstrations with TESS data}\label{sec:casestudies}

The simulations in the preceding section support that our methodology can faithfully recover a variable source location in pixel data in idealized cases (e.g., Gaussian point spread function, no data systematics), with reported errors matching the intrinsic uncertainty of the fitting procedure. Here we apply our localization tools to actual TESS data, staring with a practical analysis that we contributed to \citet{Corsico2021}.  We then apply our tools to an ensemble of eclipsing binary systems that enables us to define a model for extrinsic uncertainty and pointing systematics that we incorporate into the \tl\ software.

\subsection{Pulsations and Eclipses in the Light Curve of the GW Vir Star RX\,J2117.1+3412}\label{sec:Corsico}

\begin{figure*} 
	\begin{center}
	\includegraphics[width=1.9\columnwidth]{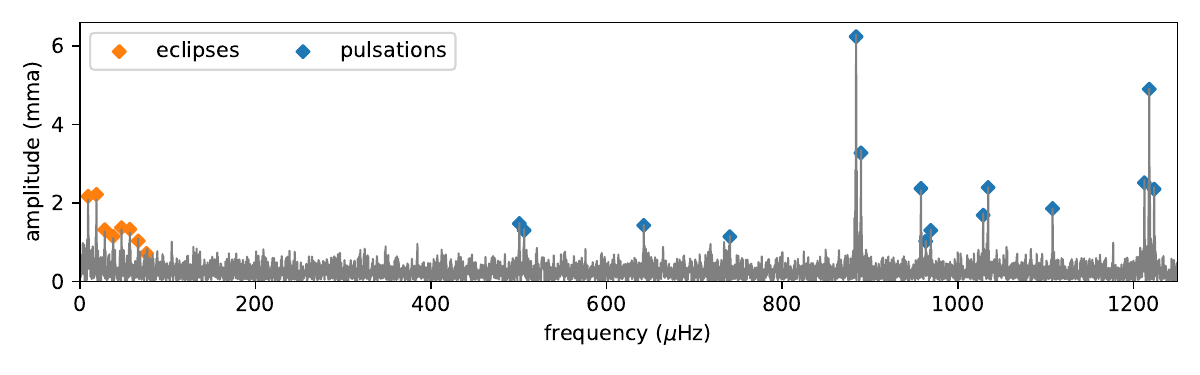}\\
	
	\includegraphics[width=.85\columnwidth,trim=1in 1in 1in 1in, clip]{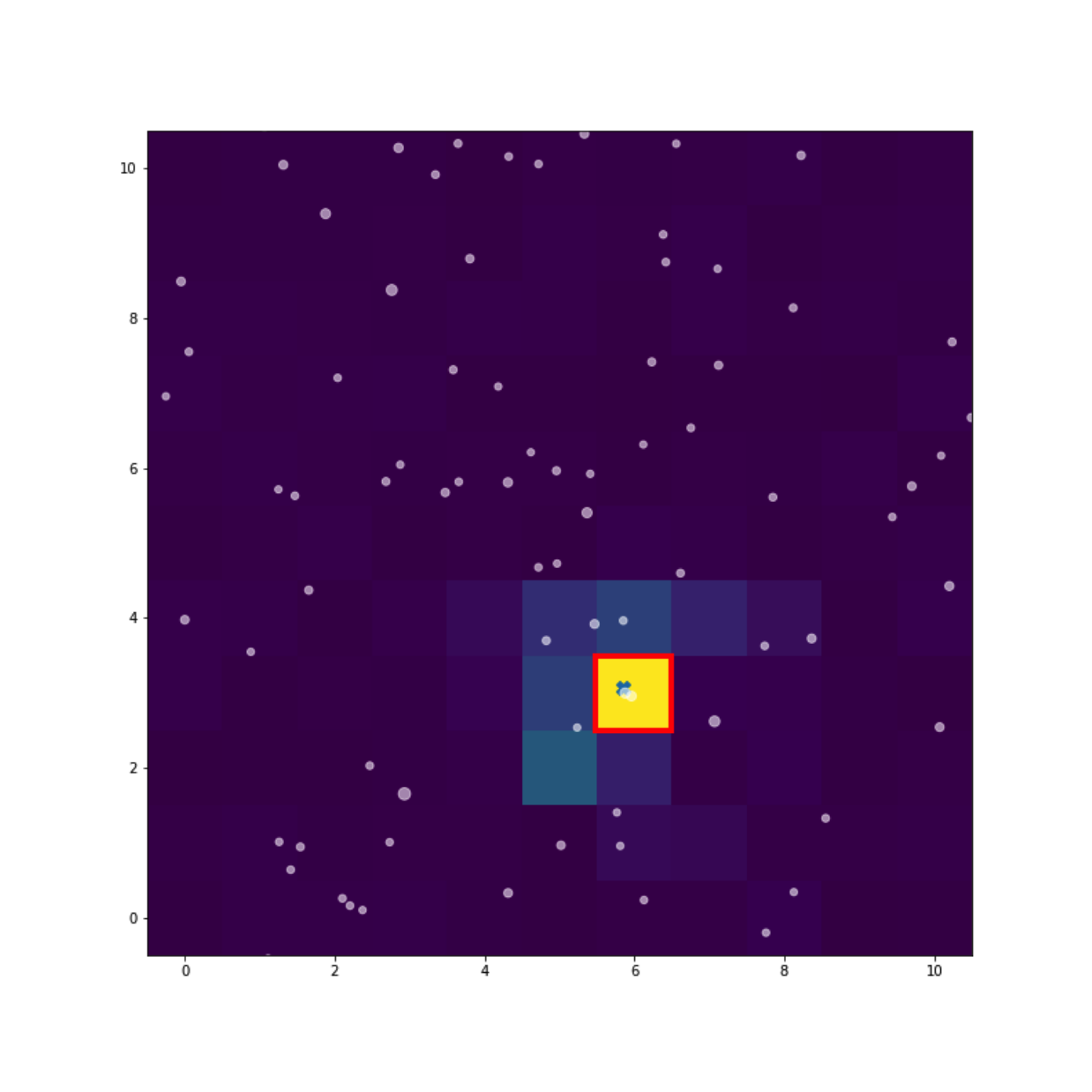}
	\includegraphics[width=.85\columnwidth,trim=1in 1in 1in 1in, clip]{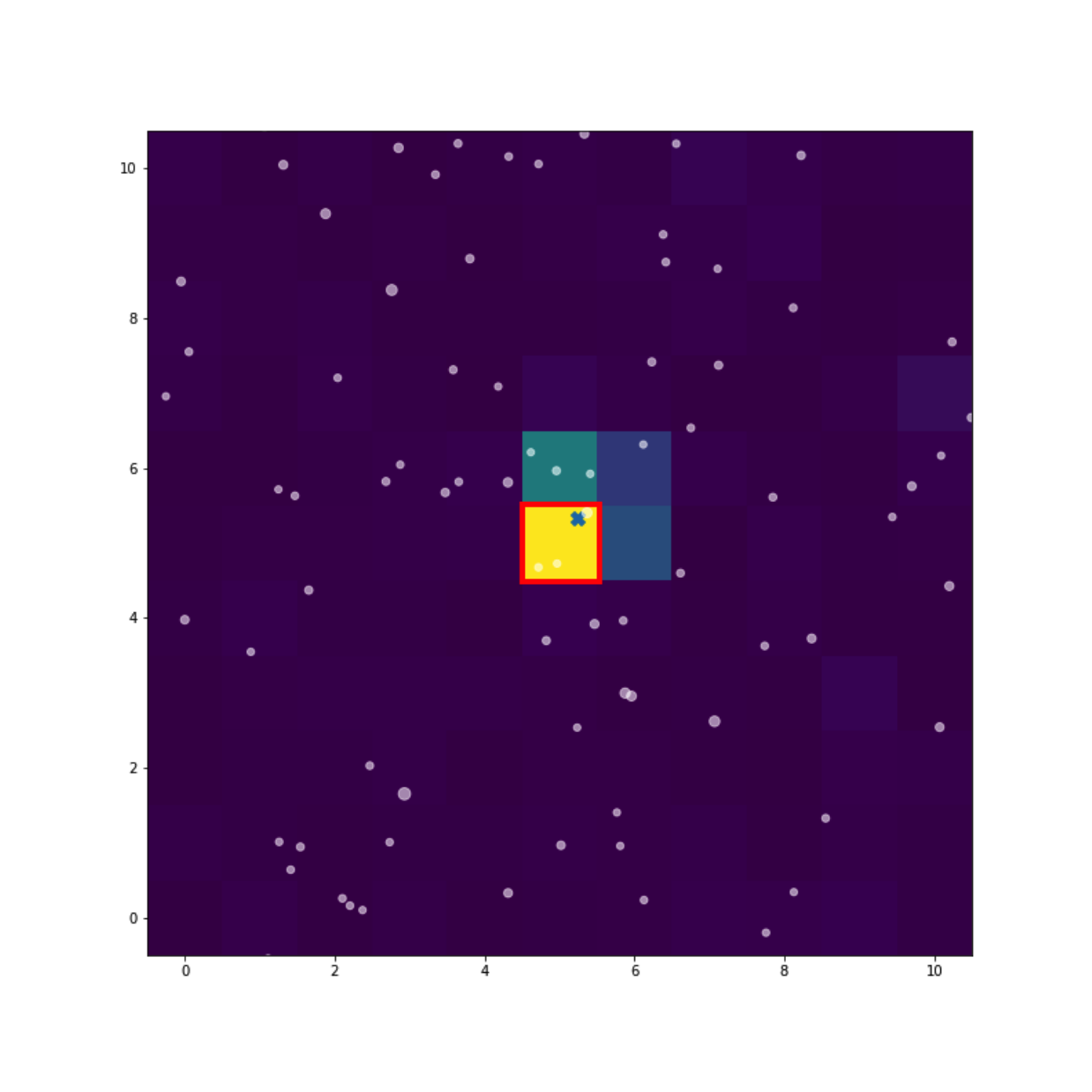}\\
	
	\includegraphics[width=1.9\columnwidth]{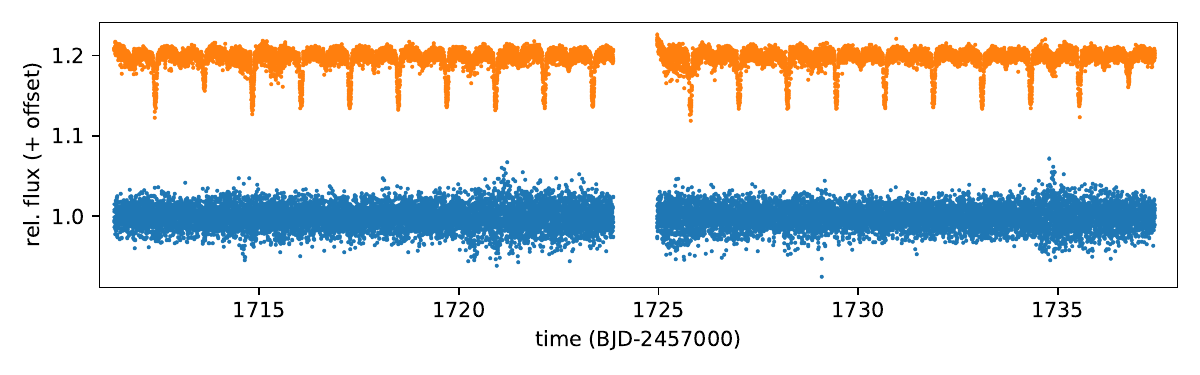}
\end{center}
	\caption{{\sc Top}: Periodogram of RX\,J2117.1+3412. Low-frequency signals of an eclipsing binary (orange) and high-frequency signals from stellar pulsations (blue) fit with {\tt Pyriod} are marked.  {\sc Middle}: Heatmaps for low- (left) and high-frequency (right) signals. These two different types of signal clearly originate from different positions on the sky. These heatmaps are a result of summing the individual amplitude heatmaps for each frequency and dividing by the square root of the sum of the squared error heatmaps. The grey points indicate the positions of \textit{Gaia} DR2 sources brighter than $G=18$\,mag in the field. {\sc Bottom}: Light curves extracted from the single hottest pixels in each heatmap (marked with a red border in heatmaps).  The binary eclipse signature is confirmed to be much more pronounced in the light curve extracted away from the white dwarf target (orange), compared to the light curve extracted at the position of the white dwarf (blue).} 
	\label{fig:rxj2117} 
\end{figure*} 

\citet{Corsico2021} carried out an analysis of 2-min cadence TESS observations of six GW Vir type pulsating white dwarfs. The light curve of one of these targets, RX\,J2117.1+3412 (TIC~117070953, Sector 15), showed evidence of both eclipses and pulsations. This implied an exciting interpretation that this hot pulsating white dwarf in a young planetary nebula could also be part of an eclipsing binary system. Due to the prevalence of source blending in TESS data, this hypothesis must be rigorously tested, and \citet{Corsico2021} were able to reject this interpretation with the use of \tl.

The Lomb-Scargle periodogram for this light curve in the region of significant variability signals is displayed in the top panel of Figure~\ref{fig:rxj2117}.  In addition to the fifteen pulsation signatures in the frequency range 500--1225\,$\mu$Hz, there appears a sequence of low-frequency harmonics at multiples of 9.511\,$\mu$Hz, indicative of binary eclipses. Because the high- and low-frequency signals appear to be associated with different physical processes, we test each set of signals independently for consistency between the concentration of power on the sky and the location of the GW Vir target. The two sets of signals that we fit with the pre-whitening frequency analysis software {\tt Pyriod}\footnote{\url{github.com/keatonb/Pyriod}} are marked with different colors in the top panel of Figure~\ref{fig:rxj2117}, and the heatmaps for each are displayed in the middle panels. These heatmaps are a result of summing the individual amplitude heatmaps for each frequency and dividing by the root sum squared of the error heatmaps.  While the power associated with pulsations is concentrated in the pixels surrounding the white dwarf RX\,J2117.1+3412, the signals from binary eclipses are significantly offset from our target source. 

Comparing to \textit{Gaia} source locations, \tl\ returns a relative likelihood that the pulsation signals originate from the white dwarf of 97.5\%. The origin of the eclipse signals is consistent with two \textit{Gaia} sources, 63.2\% with {\tt source\_id} 1855294415817908480 and 36.7\% with 1855294415817907840.
To confirm this analysis, we extract light curves from the single ``hottest'' pixel associated with each set of signals, normalize and remove long-term trends with {\tt lightkurve}, and display these in the bottom panel of Figure~\ref{fig:rxj2117}. This ``hottest'' pixel aperture can be accessed as a \tl\ class variable called {\tt maxsignal\_aperture}.  The binary eclipses are far more pronounced in the light curve extracted farther from the white dwarf target where the binary signal amplitudes appear largest, confirming that the eclipsing binary is a different contaminating source in the field.  RX\,J2117.1+3412 was dismissed as an eclipsing binary by \citet{Corsico2021} on the basis of this analysis.

The full code needed to localize the eclipse and pulsation signals in the TIC\,117070953 TPF is provided in Appendix~\ref{app:code}. Besides measuring the frequencies of variability from a periodogram, which we assume has been done already, it only takes one line of code to download the TPF data with {\tt lightkurve}, and one line of code to localize the source with \tl. The localization step takes roughly one minute on a typical consumer laptop.

\subsection{Eclipsing binary systems}\label{sec:binaries}

\citet{Prsa2022} derived a catalog of 4584 eclipsing binaries in the first two years of short-cadence TESS data.\footnote{\href{http://tessebs.villanova.edu/}{tessebs.villanova.edu}} We run \tl\ on the 120-s cadence light curves for these targets to test its performance. Starting with the orbital periods tabulated by Pr\v sa et al.,\ we use {\tt Pyriod} to automatically determine the frequencies of greatest power in the periodograms of the 3470 eclipsing binaries with orbital periods shorter than five days. The periodogram of an eclipsing binary light curve typically consists of many harmonics of the orbital frequency to reproduce the eclipse profiles, as seen in the previous example. We iteratively determine the highest periodogram value at a multiple of the orbital frequency, find a corresponding best-fit sinusoid to the light curve, update the orbital frequency estimate, and then look for the next highest harmonic.  The first signal is required to exceed seven times the median value across the periodogram. We repeat this process on the residuals of the fit to include up to the 20th harmonic (or up to the Nyquist frequency), or until no additional harmonics exceed five times the median periodogram value.  Many of these targets were observed in multiple TESS sectors, and we end up with a total of 9774 binary frequency solutions through Sector 42 (allowing for a couple hundred failed downloads), from which we can compare our fitted source location to the location of the targeted star.

We run \tl\ with these targets and frequency lists, using pipeline apertures and the {\tt auto\_pca} function to estimate the best number of PCA components (up to three) that are safe to use without potentially removing the signal of interest. For 67\% of these eclipsing binaries, {\tt auto\_pca} preferred to not perform any PCA correction, while it fit and removed the strongest one, two, and three principal components in 15\%, 7\%, and 11\% of cases, respectively.

\begin{figure}
    \centering
    \includegraphics[width=.95\columnwidth]{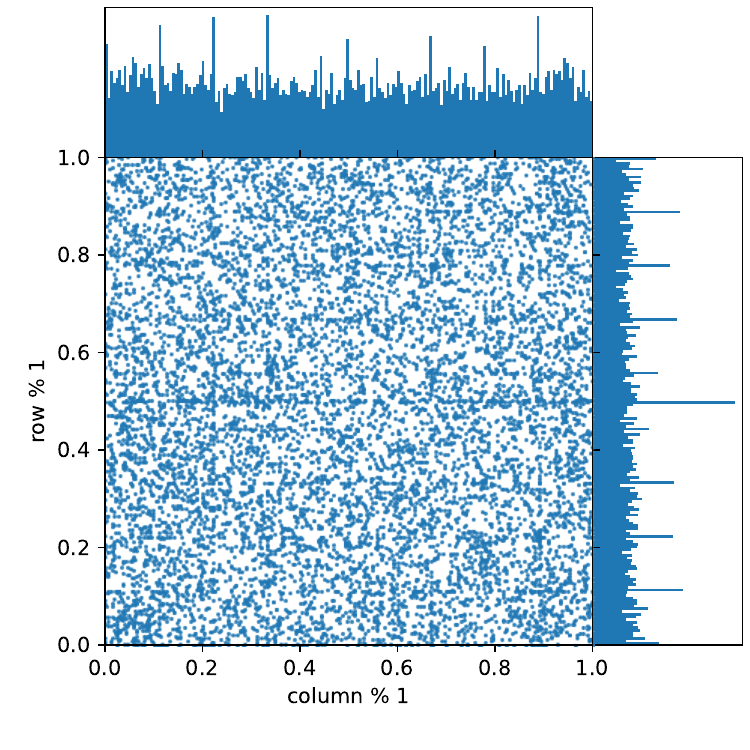}
    \caption{Sub-pixel best-fit locations for 9774 binary stars. Top and right panels display marginalized histograms. Excesses at every ninth of a pixel, as well as at the edge of pixels (0.5) indicate that the procedure of fitting a PRF model to data prefers these locations. }
    \label{fig:binsubpixel}
\end{figure}

\tl\ fits a separate height of the PRF model to the amplitude distribution for each provided frequency. Similar to source detection in an astronomical image, we should apply a significance test to determine if that signal is present over the background of noise. We require that the height of at least one component exceed five times its uncertainty to be considered a significant detection, discarding results for only 145 light curves. 
There is some structure in the resulting best-fit source locations, suggesting that there are systematics present in results from actual TESS data that did not appear in our idealized simulations from Section~\ref{sec:simulations}. Figure~\ref{fig:binsubpixel} displays the decimal portions of the source locations returned by \tl.  There are excesses every ninth of a pixel, which correspond to the locations with modelled PRFs that we interpolate between (see Appendix~\ref{app:tessprf}). There is also an excess at 0.5, corresponding to pixel edges \citep[pixel centers are defined to have integer positions; ][]{fitswcs}. It appears to be a feature of fitting with the PRF models that least-squares solutions are more likely to be found at these locations, which is an extrinsic error that will increase the fitting residuals beyond what would be expected from the formal fitting uncertainties.

To assess the effect of extrinsic error on the results, we further restrict our analysis to the fit results that have formal intrinsic fitting errors smaller than 0.05\,pix in both dimensions. The median intrinsic uncertainties are smaller than 0.01\,pix in each direction, so this rejects only the most imprecise 6\% of results in the tail of the fit distribution.

\begin{figure}
    \centering
    \includegraphics[width=.95\columnwidth]{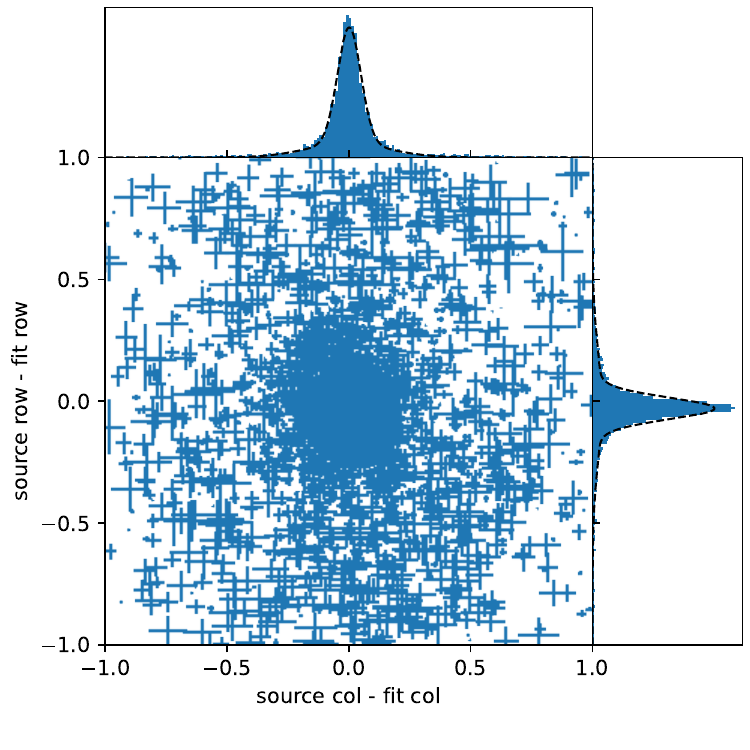}
    \caption{Residuals between the fit locations and the known locations of the targeted stars. Top and right panels display marginalized histograms, and the dashed lines show the marginalized two-component Gaussian mixture model that we adopt as an extrinsic error model in Section~\ref{sec:errormodel}}
    \label{fig:binprecision}
\end{figure}

The residuals between the fit locations and the known locations of the targeted stars are displayed in Figure~\ref{fig:binprecision}. The bulk of the results appear to be distributed about the targeted source locations. In some cases, the eclipsing binary is not the targeted source, and we would expect to fit an offset source location. Any bright star within a pixel of the target will contribute light to the aperture, so we expect this background of contaminant sources to be uniform within a pixel of the target. We use the {\tt pyGMMis} package \citep{pygmmis} to fit a two-component Gaussian mixture model (GMM) to the residuals, plus a uniform background model. The GMM distributions are indicated in Figure~\ref{fig:binprecision}, where they appear to match the marginalized histograms well. The locations, covariance matrices, and heights for the two best-fit two-dimensional Gaussians are in Table~\ref{tab:GMM}.  The uniform background model represents 11.4\% of the variable source density within one pixel of the targeted sources, and accounts for fits to sources other than the targeted stars or any spurious results.  Both Gaussian components are wider than expected from the formal fitting uncertainties, capturing the combined intrinsic and extrinsic errors.  The best-fit row locations are also notably greater than the targeted source locations according to WCS header info by $\approx 0.025$\,pix, indicating a potential systematic error in the TESS pointing model of $\approx 0.5\arcsec$.

    The header for each TPF file includes a {\sc crowdsap} keyword that estimates the fraction of flux in the photometric aperture from the targeted star. Since {\sc crowdsap} reports how contaminated a target aperture is, we might expect that variability in light curves with {\sc crowdsap} close to 1.0 should nearly always be localized to the target star.
    We find, however, that the distance between the targeted star and localization result does not have a strong apparent trend with {\sc crowdsap} value.  From inspecting our fit results, it is apparent that signals coming from sources centered well outside the photometric aperture can be present in the extracted light curves even for targets with {\sc crowdsap} $> 0.9$. This contaminating sources will populate the uniform background component of our GMM.
    
\begin{figure*}
    	\begin{center}
        	\includegraphics[width=.65\columnwidth]{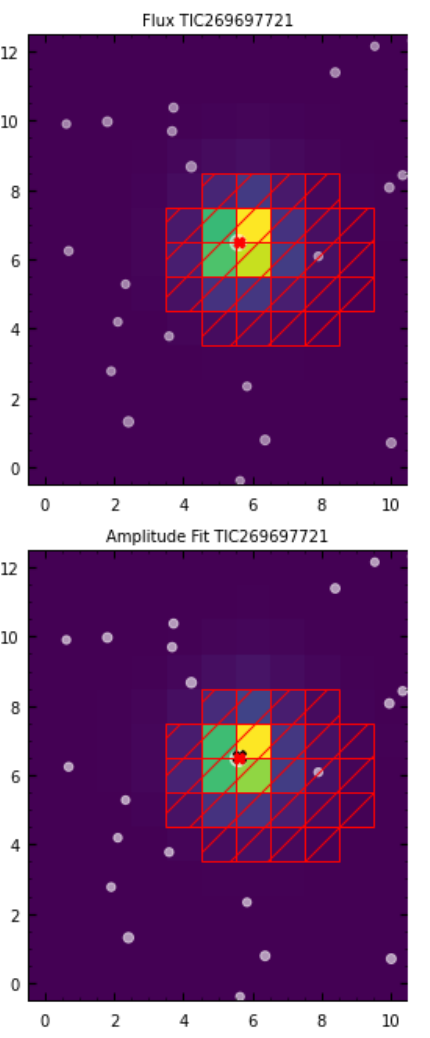}
        	\includegraphics[width=.71\columnwidth,trim=0 -.6cm 0 0]{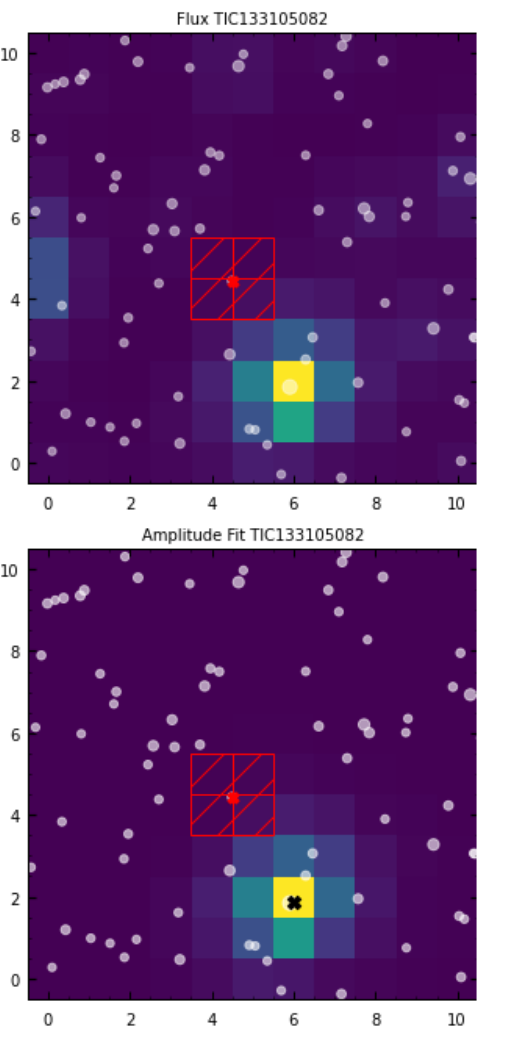}
        	\includegraphics[width=.71\columnwidth,trim=0 -.5cm 0 0]{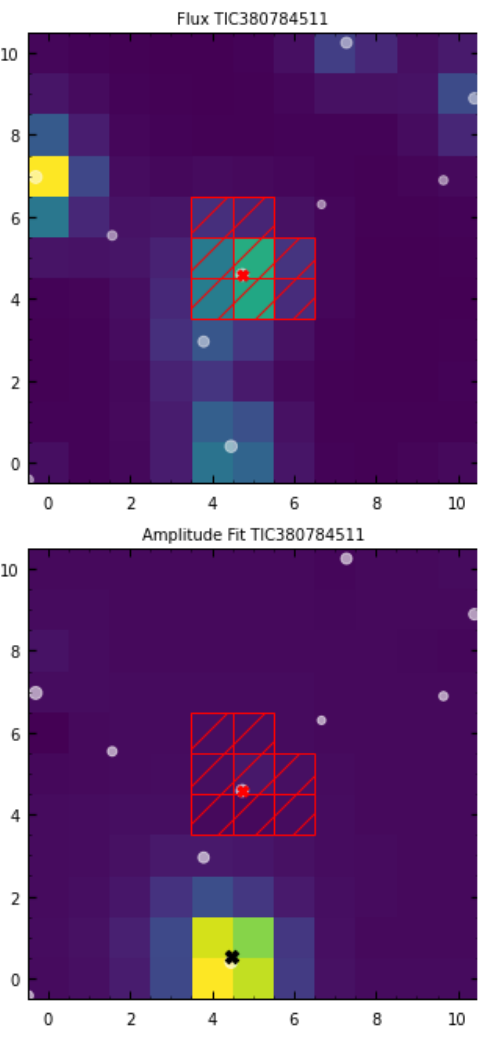}
        	
        \end{center}
    	\caption{Flux (top) and best-fit amplitude (bottom) heatmap plots for three eclipsing binaries observed by TESS \citep{Prsa2022}. From left to right, these data are for targets TIC 269697721 (Sector \#24), TIC 133105082 (Sector \#7), and TIC 380784511 (Sector \#17).  In all images, the target star is marked by a red ``X'', the localized signal is marked with a black ``X'', the target aperture is boxed in red, and the stars in the TPF are marked in grey. {\sc Left}: Example of a \tl\ fit where the target star was found to be the star exhibiting the binary eclipsing signal. This is the type of fit that is captured in the Gaussian statistics of our GMM model. {\sc Middle}: \tl\ fit where the target star has a low {\sc crowdsap}, and is not the star found to be exhibiting the binary signal. {\sc Right}: \tl\ fit where the target star has a high {\sc crowdsap}, and is not the star found to be exhibiting the binary signal.} 
    	\label{fig:fluxandampfits} 
    \end{figure*}

    In Figure~\ref{fig:fluxandampfits} we provide an example for each of the three most common situations where \tl\ succeeds in fitting the binary system. From left to right we have a high {\sc crowdsap} of 0.997 that is localized to the targeted star, a low {\sc crowdsap} of 0.114 that fits well to a contaminant star location, and a high {\sc crowdsap} of 0.921 that is also localized to a contaminant star. The top images show the flux in the TPF while the bottom images display the amplitude fits for the eclipsing frequency with the greatest signal to noise and star fit location reported by \tl. The best-fit amplitude distributions successfully isolate the flux from the single variable source of interest, making the flux from other sources apparently disappear compared to the image flux distributions.

\begin{table}
	\centering
	\caption{Extrinsic error model Gaussian mixture components (source - fit).}
	\label{tab:GMM}
	\begin{tabular}{ccc} 
		\hline
		Location & Covariance & Height \\
		(column, row) & (pix$^2$) & \\
		\hline
		(0.0018, -0.0296) &  
$\begin{bmatrix}
 0.00231 &  -0.00003\\
-0.00003 &  0.00194
\end{bmatrix}$
		& 0.626 \\
		(-0.0054, -0.0202) &
$\begin{bmatrix}
0.0282 & -0.0024 \\
-0.0024 &  0.0515
\end{bmatrix}$ & 0.259  \\
 \hline
    \end{tabular}
\end{table}

\subsubsection{Extrinsic Error Model}\label{sec:errormodel}

The GMM fit to the eclipsing binary location residuals provides a sensible empirical model for the extrinsic error on source locations fit by \tl.  Both components have scale factors significantly larger than the median intrinsic uncertainties of $<0.01$\,pix in each direction, so formal fitting errors cannot account for the residuals. The small intrinsic errors in the data will slightly increase the scale of the GMM over what would be expected from purely extrinsic errors, so using this model as a proxy for the extrinsic errors is a conservative choice that errs on the side of overestimating uncertainty.

\tl\ takes a {\tt pyGMMis} object as an extrinsic error model, with the model fit to the eclipsing binary position residuals as the default. The intrinsic error from the PRF fit given by the covariance matrix is added to the covariance matrices of the GMM model, in effect convolving the extrinsic error model by a Gaussian representation of the fit uncertainties.  When evaluating the relative likelihood of nearby \textit{Gaia} sources being the origin of observed variability, the GMM is evaluated at each source location, and the total likelihood of the variability coming from any nearby source is normalized to one, assuming that the variability originates from one of the known sources. An upper magnitude limit can be provided for \textit{Gaia} sources to include in this calculation (default: $G\leq18$\,mag), which is especially useful if the amplitude of variability rules out faint sources (e.g., an eclipse cannot block more than the total flux from a star).

\section{Discussion and Conclusions}\label{sec:conclusions}

We have demonstrated a method that can reliably localize the sources of observed variability in TESS data. The amplitudes of sinusoidal variability fit to the unnormalized light curves extracted from each pixel in a TPF are proportional to the fraction of light from the variable source captured by each pixel. A PRF model of how light from a point source is distributed on the detector can be fit to the pixel-by-pixel signal amplitudes to localize the source of variability in the image. This can be converted to a location on the sky or compared to known positions of potential sources in the images.

The localization procedure is implemented in an open-source Python package \tl, available at \href{https://github.com/Higgins00/TESS-Localize}{github.com/Higgins00/TESS-Localize}. The code requires two inputs for localization: TPF data passed as a {\tt Lightkurve} object, and the frequencies of observed variability measured from periodogram analysis. \tl\ also takes the additional optional user input: whether to make the fit location to \textit{Gaia} sources; the maximum \textit{Gaia} magnitude to consider; the unit of the provided frequencies; the number of principal components to detrend the light curves with; the aperture to use to extract the light curve for phase fitting; and a mask indicating pixels to exclude from the analysis.
The package returns a best-fit pixel position, sky coordinates, and relative probabilities of this position corresponding to known \textit{Gaia} sources.
The few lines of code needed to localize the sources of variability considered in Section~\ref{sec:Corsico} are provided in Appendix~\ref{app:code}.

Most TESS light curves represent the blended light from multiple sources due to the exceptionally instrument large plate scale of 21\arcsec\,pix$^{-1}$.  This results in a significant risk of misattributing observed variability to the wrong source.  The localization procedure developed here mitigates this risk, and we encourage that every analysis of TESS light curves include tests to confirm the source of variability.  An example analysis from \citet{Corsico2021} was detailed in Section~\ref{sec:Corsico}, where signals from pulsations and eclipses in the TPF of TIC\,117070953 were shown to come from two different sources.  \citet{Mullally2022} demonstrated a good use of \tl\ when using TESS data to vet the suitability of proposed spectrophotometric standard stars for James Webb Space Telescope instrument calibration.

The main underlying assumption of our method that must be satisfied in order to obtain reliable results is that only one source in the TPF is significantly variable at the provided frequencies. It is unlikely for multiple stars in a TPF to be variable at the same frequency by chance due to the high frequency resolution of 0.43\,$\mu$Hz for a 27-day TESS sector. This is more likely to be a problem if another source in the TPF exhibits broadband variability that spans many frequency bins.  If more than one source is variable at a provided frequency, the fitted amplitudes in the corresponding heatmap will not be distributed like the PRF. This will also be the case for bright sources that saturate the TESS pixels ($T_{\rm mag} \lesssim 6$).

Crosstalk is a phenomenon caused by electronic signals from adjacent CCD readout electronics circuits superimposing on each other and the nature of this effect can be positive or negative \citep{2018Instrumenthandbook}. Ghost images occur when point sources within and near a camera FOV are internally reflected casting sometimes spatially large images into the TPF \citep{2018Instrumenthandbook}. In the case where there is a bright variable star exhibiting crosstalk across the amplifiers, or if the variable star is a ghost from a bright star outside the field of view of the detector our software will fit the location of the variability to the ghost images or the locations of the crosstalk. In most cases these locations will not correspond to a Gaia source location. \tl\ assumes that the analyzed signals are not caused by these anomalies.

If the assumptions underlying the method are not satisfied, results obtained by \tl\ will be of poor quality. 
We recommend inspecting the following aspects of the fit results to ensure reliable localizations:
\begin{enumerate}
    \item The localization detection should be significant. We recommend that at least one of the ``height'' fit parameters for a signal amplitude exceed five times its uncertainty for the localization to be considered significantly above the noise floor. Insignificant detections will be localized to spurious positions. Individual signals that are not significant cannot be reliably associated with the fit location.
    
    \item The amplitude heatmap should be distributed like a PRF. Fitting the PRF model is the final step of localization, and the best-fit model should closely resemble the amplitude heatmaps, given their uncertainties. \tl\ provides the {\tt plot} function to display, for every input frequency, the pixel-by-pixel values for amplitude, amplitude error, signal-to-noise ratio, best-fit model, and amplitude - model residuals. If the model does not resemble the amplitude heatmap, it could indicate a problem with the localization for that target that needs to be diagnosed. Inspecting the fit residuals can be particularly illuminating. Potentially the assumption that frequencies of variability are not shared between multiple source in the TPF has been violated. The reduced Chi-square ($\chi^2_{\rm red}$) fit statistic is a measure of agreement between model and data within the amplitude fit uncertainties that could be considered a measurement of fit quality. While $\chi^2_{\rm red}$ values may be useful for flagging potential issues, we find that reliable results can be obtained for $\chi^2_{\rm red} > 1000$ in cases, and that visual inspection of the fit is most useful for understanding the quality of results.
    
    \item The light curve fit should be in phase with the intended time series signals. The best-fit sum-of-sinusoid model to the light curve extracted from the pixel containing the most model flux is displayed with the {\tt plot\_lc} function. The model should fit well to observable variability at the provided frequencies. A poor fit to the intended signal could indicate that the light curve signal-to-noise is not sufficiently high in the provided aperture to achieve a good phase fit, or that the fit was dominated by noise systematics that you could attempt to remove.
    
    \item If matching the localization of stellar variability to Gaia sources, at least one consistent source is expected. The reported p-values estimate the fraction of fit locations that would have been less likely than the actual fit location under the hypothesis that each source in the field is the variable source. This calculation considers the probability density distribution to be the convolution of intrinsic and extrinsic error models. A suitable lower threshold for continuing to consider Gaia stars as potential variability sources will depend on your experiment. For example, we expect 5\% of localizations with $p=0.05$ to not originate from the corresponding stars.
\end{enumerate}

Be aware that even with a good fit, there may be remaining confusion amongst Gaia sources. It is possible that a reliable localization will be consistent with multiple sources on the sky. The relative likelihood of each source being the origin of variability based on relative position is given by the ``relative likelihood'' column of the star fit results.

There are a couple of things you can try to improve the quality of \tl\ results if you identify one of the above issues:
\begin{enumerate}
    \item Try using different PCA components in the detrending step. The aim is to remove systematic trends in the light curves without removing the variability of interest. It is encouraged to inspect the light curves and periodograms of PCA trends determined from pixels outside the provided aperture to decide which should be used in the analysis.
    
    \item It is important that the signal you wish to localize is strong within the provided aperture (pipeline aperture by default). Signal phases are fixed by \tl\ to the values that fit best to the light curve extracted from the aperture, so the quality of results are limited to the precision of this initial fit. Fit uncertainties on phase $< 0.1$\,radians is a good proxy for a localizable signal (Section~\ref{sec:case1}).  It may be the case that most power at the given frequencies is located outside the aperture, and that a better fit could be obtained by providing a different aperture.  The {\tt aperture="auto"} option attempts to automatically choose a best aperture where the Lomb-Scargle periodogram values at the input frequencies are highest. A more precise localization may be obtained by using the results of an initial fit to choose a new aperture that better captures the signal of interest. If most of the light from the variable source is located off the edge of the TPF, a more suitable set of pixels could be obtained from the FFI with {\tt TESSCut} \citep{TESSCut}, though only at the long FFI cadence which may be insensitive to high-frequency variability.
    
    \item It may be the case that the localization is being attempted on a set of signal frequencies that originate from multiple sources. It is advised to test that subsets of signal frequencies appear to originate from a consistent location. Keep in mind that signals that are not fit with significant ``heights'' cannot be reliably associated with the fit location unless they can be physically associated with other signals that are significantly detected (e.g., signal harmonics).
    
    \item Finally, the flux recorded by some individual pixels could be of poor quality and disrupt the fitting procedure (e.g, saturated, contaminated by moving asteroids, otherwise dominated by severe noise). A pixel mask can be provided to exclude specified pixels from analyses in these situations.
\end{enumerate}

\section*{Acknowledgements}

We thank the anonymous referee for their feedback that helped to improve this paper. K.J.B.\ is supported by the National Science Foundation under Award AST-1903828. 
We thank Jon Jenkins, Rebekah Hounsell, Roland Vanderspek, Michael Fausnaugh, Scott Fleming, Susan Mullally, and Andrej {Pr{\v{s}}a} for helpful discussions.
This paper includes data collected with the TESS mission, obtained from the MAST data archive at the Space Telescope Science Institute (STScI). Funding for the TESS mission is provided by the NASA Explorer Program. STScI is operated by the Association of Universities for Research in Astronomy, Inc., under NASA contract NAS 5–26555. This research made use of Lightkurve, a Python package for Kepler and TESS data analysis \citep{lightkurve}.
This work has made use of data from the European Space Agency (ESA) mission
{\it Gaia} (\url{https://www.cosmos.esa.int/gaia}), processed by the {\it Gaia}
Data Processing and Analysis Consortium (DPAC,
\url{https://www.cosmos.esa.int/web/gaia/dpac/consortium}). Funding for the DPAC
has been provided by national institutions, in particular the institutions
participating in the {\it Gaia} Multilateral Agreement.
Resources supporting this work were provided by the NASA High-End Computing (HEC) Program through the NASA Advanced Supercomputing (NAS) Division at Ames Research Center for the production of the SPOC data products.

\software{Astropy \citep{astropy:2013,astropy:2018}, astroquery \citep{2019AJ....157...98G}, 
Lightkurve \citep{lightkurve},
lmfit \citep{lmfit},
Matplotlib \citep{Hunter:2007}, pyGMMis \citep{pygmmis}
}

\bibliography{DraftPaper}{}
\bibliographystyle{aasjournal}

\appendix

\section{\tl\ Code example}\label{app:code}

The Python code below reproduces the localizations of the sources of both pulsational and eclipsing binary variability observed in the TPF for target TIC\,117070953 that was presented in Section~\ref{sec:Corsico}. 

\begin{lstlisting}[language=sh]
import TESS_Localize as tl
import lightkurve as lk
import astropy.units as u

#Binary frequencies
low_frequency_list = [9.51112996, 19.02225993, 28.53338989, 38.04451986,
                      47.55564982, 57.06677979, 66.57790975, 76.08903972]

#Pulsation frequencies
high_frequency_list = [500.559, 506.057, 642.255, 740.266, 884.017,
                       889.556, 957.817, 963.28, 969.013, 1028.729,
                       1034.356, 1107.713, 1212.297, 1217.872, 1223.429]

#Download TPF data for this target with lightkurve
tpf = lk.search_targetpixelfile('TIC117070953', sector=15, cadence=120).download()

#Localize the low frequencies (binary eclipses)
low = tl.Localize(targetpixelfile=tpf, frequencies=low_frequency_list, frequnit=u.uHz, 
                  principal_components='auto')

#Localize the high frequencies (stellar pulsations)
high = tl.Localize(targetpixelfile=tpf, frequencies=high_frequency_list, frequnit=u.uHz, 
                   principal_components='auto')
\end{lstlisting}

\section{Sampling the TESS Pixel Response Function with \texttt{TESS\_PRF}}\label{app:tessprf}

As an ancillary product of this work, we developed a Python package for interpolating the TESS Pixel Response Function (PRF) models called {\tt TESS\_PRF}, available at \href{https://github.com/keatonb/TESS_PRF}{github.com/keatonb/TESS\_PRF}.
The PRF describes how light from a point source will be distributed across pixels at different positions on the detector.  It accounts for the optical point spread function, pointing jitter during an exposure, and intra-pixel sensitivity \citep{2019JATIS...5d1507V}.
The PRF models are an important TESS data product with many potential uses, such as simulating realistic TESS data or performing PRF photometry. In this work, we used the PRF models to fit pixel-by-pixel amplitude distributions to achieve our source localization results.

The TESS PRF model files are available on MAST at \href{https://heasarc.gsfc.nasa.gov/docs/tess/observing-technical.html#point-spread-function}{heasarc.gsfc.nasa.gov/docs/tess/observing-technical.html\#point-spread-function}. 
They were created by the TESS Science Processing Operations Center \citep{2016SPIE.9913E..3EJ} following the methods developed for the \textit{Kepler} spacecraft described in \citet{Bryson2010}.
Different sets of models are applicable to TESS Sectors 1-3 and Sectors 4 and later. For each camera and CCD, PRF model files are available for a grid of pixels separated by 512 columns or rows. Each of these files contains 81 PRF models depending on where the flux source is located within a given pixel.  These are sampled every ninth of a pixel in each direction, starting from 1/18th of a pixel from the edge, and with the central model corresponding to the center of the pixel.  These models are interleaved so that the PRF values at one of these subpixel locations are given by every ninth array value. The esoteric format of these PRF files motivated our creation of the {\tt TESS\_PRF} package.

{\tt TESS\_PRF} was built for local analysis of TESS pixel subregions, such as a TPF, so that the PRF can be assumed not to change appreciably across the image.  As a first step, {\tt TESS\_PRF} interpolates between the available PRF files to a given pixel location, which should be near the center of the TPF you wish to analyze.  It will access the relevant files from MAST by default, or it can be pointed to a local directory containing these files for faster or offline analysis.  After this first interpolation, a second interpolation is performed to position the model at a given subpixel location within the TPF with the {\tt locate} function.  See package documentation for detailed usage instructions.


\end{document}